\documentclass[12pt]{article}
\usepackage{amssymb,amscd,array}
\catcode `\@=11
\@addtoreset{equation}{section}

\newtheorem{thm}{Theorem}[section]

\newtheorem{lemma}[thm]{Lemma}
\newtheorem{prop}[thm]{Proposition}

\def\qed{\blacksquare}
\newcommand{\be}{\begin{equation}}
\newcommand{\ee}{\end{equation}}
\newcommand{\bea}{\begin{eqnarray}}
\newcommand{\eea}{\end{eqnarray}}
\newcommand{\R}{\mathbb{R}}

\begin{document}
\begin{titlepage}

\begin{center}
{\bf \Large{A New Formula for the Gauge Charge \\}}
\end{center}
\vskip 1.0truecm
\centerline{D. R. Grigore, 
\footnote{e-mail: grigore@theory.nipne.ro}}
\vskip5mm
\centerline{Department of Theoretical Physics,}
\centerline{Institute for Physics and Nuclear Engineering ``Horia Hulubei"}
\centerline{Bucharest-M\u agurele, P. O. Box MG 6, ROM\^ANIA}

\vskip 2cm
\bigskip \nopagebreak
\vskip 1cm
\begin{abstract}
\noindent
We present a new formula for the gauge charge in the causal formalism for the QCD case.
\end{abstract}

\end{titlepage}

\section{Introduction}

The most natural way to arrive at the Bogoliubov axioms of perturbative quantum field theory (pQFT) is by analogy with non-relativistic 
quantum mechanics \cite{Gl}, \cite{H}, \cite{D}, \cite{DF}. So we start from Bogoliubov axioms \cite{BS}, \cite{EG}, \cite{Sc1}, \cite{Sc2} 
as presented in \cite{cohomology}, \cite{algebra}. 
The Bogoliubov axioms express essentially some properties of the scattering matrix understood as a formal perturbation series with the 
``coefficients" the chronological products: 
(1) (skew)symmetry property; 
(2) Poincar\'e invariance; 
(3) causality; 
(4) unitarity; 
(5) the ``initial condition" which says that the first order chronological products are Wick monomials.
So we need some basic notions on free fields and Wick monomials which will be presented in Section \ref{wick prod} 
also following \cite{algebra}.  One can supplement these axioms by requiring 
(6) power counting; 
(7) Wick expansion property. 
It is a highly non-trivial problem to find solutions for the Bogoliubov axioms, even in the simplest case of a real scalar field. 

There are, at least to our knowledge, three rigorous ways to do that; for completeness we remind them following \cite{ano-free}:
(a) {\it Hepp axioms} \cite{H}, \cite{Z-J};
(b) {\it Polchinski flow equations} \cite{P}, \cite{S};
(c) {\it the causal approach} due to Epstein and Glaser \cite{EG}, \cite{Gl} which we prefer. 
It is a recursive procedure and reduces the induction procedure to a distribution splitting of some distributions with causal support.  
In an equivalent way, one can reduce the induction procedure to the process of extension of distributions \cite{PS}. 

An equivalent point of view uses retarded products \cite{St1} instead of chronological products. For gauge models one has to deal with 
non-physical fields (the so-called ghost fields) and impose a supplementary axiom (8) namely  gauge invariance, which guarantees that the 
physical states are left invariant by the chronological products.

In this paper we will use consider the causal approach and, using some properties of the Wick products, we will prove a new formula for the 
gauge charge.  
\newpage

\section{Wick Products\label{wick prod}}

We follow the formalism from \cite{algebra}. We consider a classical field theory on the Minkowski space
$
{\cal M} \simeq \R^{4}
$
(with variables
$
x^{\mu}, \mu = 0,\dots,3
$
and the metric $\eta$ with 
$
diag(\eta) = (1,-1,-1,-1)
$)
described by the Grassmann manifold 
$
\Xi_{0}
$
with variables
$
\xi_{a}, a \in {\cal A}
$
(here ${\cal A}$ is some index set) and the associated jet extension
$
J^{r}({\cal M}, \Xi_{0}),~r \geq 1
$
with variables 
$
x^{\mu},~\xi_{a;\mu_{1},\dots,\mu_{n}},~n = 0,\dots,r;
$
we denote generically by
$
\xi_{p}, p \in P
$
the variables corresponding to classical fields and their formal derivatives and by
$
\Xi_{r}
$
the linear space generated by them. The variables from
$
\Xi_{r}
$
generate the algebra
$
{\rm Alg}(\Xi_{r})
$
of polynomials.

In classical field theory the jet-bundle extensions do verify Euler-Lagrange equations. To write them we need the formal derivatives 
defined by
\be
d_{\nu}\phi_{\{\mu_{1},\dots,\mu_{r}\}} \equiv \phi_{\{\nu,\mu_{1},\dots,\mu_{r}\}}.
\ee

We suppose that in the algebra 
$
{\rm Alg}(\Xi_{r})
$
generated by the variables 
$
\xi_{p}
$
there is a natural conjugation
$
A \rightarrow A^{\dagger}.
$
If $A$ is some monomial in these variables, there is a canonical way to associate to $A$ a Wick 
monomial: we associate to every classical field
$
\xi_{a}, a \in {\cal A}
$
a quantum free field denoted by
$
\xi^{\rm quant}_{a}(x), a \in {\cal A}
$
and determined by the $2$-point function
\be
<\Omega, \xi^{\rm quant}_{a}(x), \xi^{\rm quant}_{b}(y) \Omega> = - i~D^{(+)}(\xi_{a}(x),\xi_{b}(y))\times {\bf 1}.
\label{2-point}
\ee
Here 
\be
D_{ab}(x - y) \equiv D(\xi_{a}(x),\xi_{b}(y))
\ee
is the causal Pauli-Jordan distribution associated to the two fields; it is (up to some numerical factors) a polynomial
in the derivatives applied to the Pauli-Jordan distribution. We understand by 
$
D^{(\pm)}_{ab}(x)
$
the positive and negative parts of
$
D_{ab}(x)
$.
The $n$-point functions for
$
n \geq 3
$
are obtained assuming that the truncated Wightman functions are null: see \cite{BLOT}, relations (8.74) and (8.75) and proposition 8.8
from there. The definition of these truncated Wightman functions involves the Fermi parities
$
|\xi_{p}|
$
of the fields
$
\xi_{p}, p \in P.
$

Afterwards we define
$$
\xi^{\rm quant}_{a;\mu_{1},\dots,\mu_{n}}(x) \equiv \partial_{\mu_{1}}\dots \partial_{\mu_{n}}\xi^{\rm quant}_{a}(x), a \in {\cal A}
$$
which amounts to
\be
D( \xi_{a;\mu_{1}\dots\mu_{m}}(x), \xi_{b;\nu_{1}\dots\nu_{n}}(y) ) =
(-1)^{n}~i~\partial_{\mu_{1}}\dots \partial_{\mu_{m}}\partial_{\nu_{1}}\dots \partial_{\nu_{n}}D_{ab}(x - y )\times {\bf 1}.
\label{2-point-der}
\ee
More sophisticated ways to define the free fields involve the GNS construction. 

The free quantum fields are generating a Fock space 
$
{\cal F}
$
in the sense of the Borchers algebra: formally it is generated by states of the form
$
\xi^{\rm quant}_{a_{1}}(x_{1})\dots \xi^{\rm quant}_{a_{n}}(x_{n})\Omega
$
where 
$
\Omega
$
the vacuum state.
The scalar product in this Fock space is constructed using the $n$-point distributions and we denote by
$
{\cal F}_{0} \subset {\cal F}
$
the algebraic Fock space.

One can prove that the quantum fields are free, i.e.
they verify some free field equation; in particular every field must verify Klein Gordon equation for some mass $m$
\be
(\square + m^{2})~\xi^{\rm quant}_{a}(x) = 0
\label{KG}
\ee
and it follows that in momentum space they must have the support on the hyperboloid of mass $m$. This means that 
they can be split in two parts
$
\xi^{\rm quant (\pm)}_{a}
$
with support on the upper (resp. lower) hyperboloid of mass $m$. We convene that 
$
\xi^{\rm quant (+)}_{a} 
$
resp.
$
\xi^{\rm quant (-)}_{a} 
$
correspond to the creation (resp. annihilation) part of the quantum field. The expressions
$
\xi^{\rm quant (+)}_{p} 
$
resp.
$
\xi^{\rm quant (-)}_{p} 
$
for a generic
$
\xi_{p},~ p \in P
$
are obtained in a natural way, applying partial derivatives. For a general discussion of this method of constructing free fields, see 
ref. \cite{BLOT} - especially prop. 8.8.
The Wick monomials are leaving invariant the algebraic Fock space.
The definition for the Wick monomials is contained in the following Proposition.

\begin{prop}
The operator-valued distributions
$
N(\xi_{q_{1}}(x_{1}),\dots,\xi_{q_{n}}(x_{n}))
$
are uniquely defined by:

\be
N(\xi_{q_{1}}(x_{1}),\dots,\xi_{q_{n}}(x_{n}))\Omega = \xi_{q_{1}}^{(+)}(x_{1})\dots  \xi_{q_{n}}^{(+)}(x_{n})\Omega
\ee

\bea
[ \xi_{p}(y), N(\xi_{q_{1}}(x_{1}),\dots,\xi_{q_{n}}(x_{n})) ] = 
\nonumber\\
- i~\sum_{m=1}^{n} \prod_{l <m} (-1)^{|\xi_{p}||\xi_{q_{l}}|}~D_{pq_{m}}(y - x_{m})~N(\xi_{q_{1}}(x_{1}),\dots,\hat{m},\dots,\xi_{q_{n}}(x_{n}))
\eea

\be
N(\emptyset) = I.
\ee

The expression
$
N(\xi_{q_{1}}(x_{1}),\dots,\xi_{q_{n}}(x_{n}))
$
is (graded) symmetrical in the arguments.
\end{prop}

The expressions
$
N(\xi_{q_{1}}(x_{1}),\dots,\xi_{q_{n}}(x_{n}))
$
are called {\it Wick monomials}. There is an alternative definition based on the splitting of the fields into the creation and annihilation 
part for which we refer to \cite{algebra}.

It is a non-trivial result of Wightman and G\aa rding \cite{WG} that in
$
N(\xi_{q_{1}}(x_{1}),\dots,\xi_{q_{n}}(x_{n}))
$
one can collapse all variables into a single one and still gets an well-defined expression: if we compute the formal expression
$
N(\xi_{q_{1}}(x),\dots,\xi_{q_{n}}(x))
$
one can prove that it is well defined. So we have for every monomial $A$ in the jet variables an associated Wick monomial
$
N(A(x)).
$
For details we refer to \cite{algebra}. One can prove that 
\be
[ N(A(x)), N(B(y)) ] = 0,\quad (x - y)^{2} < 0
\ee
where by
$
[ \cdot,\cdot]
$
we mean the graded commutator. This is the most simple case of causal support property.

We can define more general expressions of the type:
$
N(A_{1}(x_{1}),\dots,A_{n}(x_{n}))
$
where we do the ``collapsing" trick in $n$ groups of variable. Now we are ready for the most general setting. We define for any monomial
$
A \in {\rm Alg}(\Xi_{r})
$
the derivation
\be
\xi \cdot A \equiv (-1)^{|\xi| |A|}~{\partial \over \partial \xi}A
\label{derivative}
\ee
for all
$
\xi \in \Xi_{r}.
$
Here 
$|A|$ 
is the Fermi parity of $A$ and we consider the left derivative in the Grassmann sense. An expression
$
E(A_{1}(x_{1}),\dots,A_{n}(x_{n}))
$
is called {\it of Wick type} iff verifies:

\bea
[ \xi_{p}(y), E(A_{1}(x_{1}),\dots,A_{n}(x_{n}))  ] = 
\nonumber\\
- i~\sum_{m=1}^{n} \prod_{l \leq m} (-1)^{|\xi_{p}||A_{l}|}~\sum_{q}~D_{pq}(y - x_{m})~
E(A_{1}(x_{1}),\dots,\xi_{q}\cdot A_{m}(x_{m}),\dots,A_{n}(x_{n}))
\label{comm-wick}
\eea

\be
E(A_{1}(x_{1}),\dots,A_{n}(x_{n}),{\bf 1}) = E(A_{1}(x_{1}),\dots,A_{n}(x_{n})) 
\ee

\be
E(1) = {\bf 1}.
\ee

The expression
$
N(A_{1}(x_{1}),\dots,A_{n}(x_{n}))
$
is of Wick type. 

Now we give some basic ideas about the Bogoliubov axioms \cite{algebra}. Suppose the monomials
$
A_{1},\dots,A_{n} \in {\rm Alg}(\Xi_{r})
$
are self-adjoint:
$
A_{j}^{\dagger} = A_{j},~\forall j = 1,\dots,n
$
and of Fermi number
$
f_{i}.
$

The chronological products
$$ 
T(A_{1}(x_{1}),\dots,A_{n}(x_{n})) \equiv T^{A_{1},\dots,A_{n}}(x_{1},\dots,x_{n}) \quad n = 1,2,\dots
$$
are some distribution-valued operators leaving invariant the algebraic Fock space and verifying a set of axioms
which we have spelled out in the Introduction. We give only {\bf Wick expansion property}: In analogy to (\ref{comm-wick}) we require
\bea
[ \xi_{p}(y), T(A_{1}(x_{1}),\dots,A_{n}(x_{n})) ]
\nonumber\\
= - i~\sum_{m=1}^{n}~\prod_{l \leq m} (-1)^{|\xi_{p}||A_{l}|}~\sum_{q}~D_{pq}(y - x_{m} )~
T(A_{1}(x_{1}),\dots,\xi_{q}\cdot A_{m}(x_{m}),\dots, A_{n}(x_{n}))
\nonumber\\
\label{wick}
\eea

Up to now, we have defined the chronological products only for self-adjoint Wick monomials 
$
W_{1},\dots,W_{n}
$
but we can extend the definition for Wick polynomials by linearity.

The construction of Epstein-Glaser is based on a recursive procedure \cite{EG}. 

We provide now a consequence of (\ref{wick}); in fact we can impose a sharper form:
\bea
[ \xi_{p}^{(\epsilon)}(y), T(A_{1}(x_{1}),\dots,A_{n}(x_{n})) ]
\nonumber\\
= - i~\sum_{m=1}^{n}~\prod_{l \leq m} (-1)^{|\xi_{p}||A_{l}|}~\sum_{q}~D_{pq}^{(-\epsilon)}(y - x_{m} )~
T(A_{1}(x_{1}),\dots,\xi_{q}\cdot A_{m}(x_{m}),\dots, A_{n}(x_{n})).
\nonumber\\
\label{wick-e}
\eea
\newpage
We have:
\begin{thm}
We suppose that the variables 
$
\xi_{p}
$
are of Bose type. Then the chronological products can be chosen such that we have:
\bea
[ : \xi_{p}(y) \xi_{q}(y):, T(A_{1}(x_{1}),\dots,A_{n}(x_{n})) ] 
\nonumber\\
= - i~\xi_{p}^{(+)}(y) \sum_{m=1}^{n} D_{qr}(y - x_{m} )~
T(A_{1}(x_{1}),\dots,\xi_{r}\cdot A_{m}(x_{m}),\dots, A_{n}(x_{n}))
\nonumber\\
- i \sum_{m=1}^{n} \sum_{r}~D_{pr}(y - x_{m} )~
T(A_{1}(x_{1}),\dots,\xi_{r}\cdot A_{m}(x_{m}),\dots, A_{n}(x_{n}))~\xi_{q}^{(-)}(y)
\nonumber\\
- i \xi_{q}^{(+)}(y) \sum_{m=1}^{n} ~\sum_{r}~D_{pr}(y - x_{m} )~
T(A_{1}(x_{1}),\dots,\xi_{r}\cdot A_{m}(x_{m}),\dots, A_{n}(x_{n}))
\nonumber\\
- i \sum_{m=1}^{n} \sum_{r}~D_{qr}(y - x_{m} )~
T(A_{1}(x_{1}),\dots,\xi_{r}\cdot A_{m}(x_{m}),\dots, A_{n}(x_{n}))~\xi_{p}^{(-)}(y)
\nonumber\\
- \sum_{k < m} ~\sum_{r,s} d_{pr,qs}(y - x_{m}, y - x_{k})~
T(A_{1}(x_{1}),\dots,\xi_{s}\cdot A_{k}(x_{k}),\dots, \xi_{r}\cdot A_{m}(x_{m}),\dots,A_{n}(x_{n}))
\nonumber\\
- \sum_{k > m} ~\sum_{r,s} d_{pr,qs}(y - x_{m},y - x_{k})~
T(A_{1}(x_{1}),\dots,\xi_{r}\cdot A_{m}(x_{m}),\dots, \xi_{s}\cdot A_{k}(x_{k}),\dots,A_{n}(x_{n}))
\nonumber\\
- \sum_{m=1}^{n} \sum_{r,s} d_{pr,qs}(y - x_{m})
(A_{1}(x_{1}),\dots,\xi_{s}\cdot\xi_{r}\cdot A_{m}(x_{m}),\dots, A_{n}(x_{n}))~~~
\label{pqT}
\eea
where in the last line we apply iteratively the product (\ref{derivative}) and we have defined
\be
d_{pr,qs}(x_{1}, x_{2}) \equiv D_{pr}^{(+)}(x_{1})~D_{qs}^{(+)}(x_{2}) - D_{pr}^{(-)}(x_{1})~D_{qs}^{(-)}(x_{2})
\ee
and
\be
d_{pr,qs}(x) \equiv D_{pr}^{(+)}(x)~D_{qs}^{(+)}(x) - D_{pr}^{(-)}(x)~D_{qs}^{(-)}(x) = d_{pr,qs}(x, x).
\ee
For the general Grassmann case one has to introduce apropriate Fermi signs. 
\label{dcomm}
\end{thm}
{\bf Proof:}
We use the definition
\be
:\xi_{p}(y) \xi_{q}(y): = \xi^{(+)}_{p}(y) \xi^{(+)}_{q}(y) + \xi^{(+)}_{p}(y) \xi^{(-)}_{q}(y)
+ \xi^{(+)}_{q}(y) \xi^{(+)}_{p}(y) + \xi^{(-)}_{p}(y) \xi^{(-)}_{q}(y)
\ee
(with apropriate Fermi signs in the general case) and we can perform the commutation with
$
T(A_{1}(x_{1}),\dots,A_{n}(x_{n})) 
$
using the well-known rule
$
[AB, C] = [A,B]C + A[B,C].
$
$\qed$
\newpage
\section{Yang-Mills Fields\label{ym}}

First, we can generalize the preceding formalism to the case when some of the scalar fields
are odd Grassmann variables. One simply insert everywhere the Fermi sign. The next generalization is to arbitrary vector and spinorial
fields. If we consider for instance the Yang-Mills interaction Lagrangian corresponding to pure QCD \cite{algebra} then the jet variables 
$
\xi_{a}, a \in \Xi
$
are
$
(v^{\mu}_{a}, u_{a}, \tilde{u}_{a}),~a = 1,\dots,r
$
where 
$
v^{\mu}_{a}
$
are Grassmann even and 
$
u_{a}, \tilde{u}_{a}
$
are Grassmann odd variables. 

The interaction Lagrangian is determined by gauge invariance. Namely we define the {\it gauge charge} operator by
\be
d_{Q} v^{\mu}_{a} = i~d^{\mu}u_{a},\qquad
d_{Q} u_{a} = 0,\qquad
d_{Q} \tilde{u}_{a} = - i~d_{\mu}v^{\mu}_{a},~a = 1,\dots,r
\ee
where 
$
d^{\mu}
$
is the formal derivative. The gauge charge operator squares to zero:
\be
d_{Q}^{2} \simeq  0
\ee
where by
$
\simeq
$
we mean, modulo the equation of motion. Now we can define the interaction Lagrangian by the relative cohomology relation:
\be
d_{Q}T(x) \simeq {\rm total~divergence}.
\ee
If we eliminate the corresponding coboundaries, then a tri-linear Lorentz covariant 
expression is uniquely given by
\bea
T = f_{abc} \left( {1\over 2}~v_{a\mu}~v_{b\nu}~F_{c}^{\nu\mu}
+ u_{a}~v_{b}^{\mu}~d_{\mu}\tilde{u}_{c}\right)
\label{Tint}
\eea
where
\be
F^{\mu\nu}_{a} \equiv d^{\mu}v^{\nu}_{a} - d^{\nu}v^{\mu}_{a}, 
\quad \forall a = 1,\dots,r
\ee 
and 
$
f_{abc}
$
are real and completely anti-symmetric. (This is the tri-linear part of the usual QCD interaction Lagrangian from classical field theory.)

Then we define the associated Fock space by the non-zero $2$-point distributions are
\bea
<\Omega, v^{\mu}_{a}(x_{1}) v^{\nu}_{b}(x_{2})\Omega> = 
i~\eta^{\mu\nu}~\delta_{ab}~D_{0}^{(+)}(x_{1} - x_{2}),
\nonumber \\
<\Omega, u_{a}(x_{1}) \tilde{u}_{b}(x_{2})\Omega> = - i~\delta_{ab}~D_{0}^{(+)}(x_{1} - x_{2}),
\nonumber\\
<\Omega, \tilde{u}_{a}(x_{1}) u_{b}(x_{2})\Omega> = i~\delta_{ab}~D_{0}^{(+)}(x_{1} - x_{2}).
\label{2-massless-vector}
\eea
and construct the associated Wick monomials. Then the expression (\ref{Tint}) gives a Wick polynomial 
$
T^{\rm quant}
$
formally the same, but: 
(a) the jet variables must be replaced by the associated quantum fields; (b) the formal derivative 
$
d^{\mu}
$
goes in the true derivative in the coordinate space; (c) Wick ordering should be done to obtain well-defined operators. We also 
have an associated {\it gauge charge} operator in the Fock space given by
\bea
~[Q, v^{\mu}_{a}] = i~\partial^{\mu}u_{a},\qquad
\{ Q, u_{a} \} = 0,\qquad
\{Q, \tilde{u}_{a}\} = - i~\partial_{\mu}v^{\mu}_{a}
\nonumber \\
Q \Omega = 0.
\label{Q-vector-null}
\eea

Then it can be proved that
$
Q^{2} = 0
$
and
\be
~[Q, T^{\rm quant}(x) ] = {\rm total~divergence}
\ee
where the equations of motion are automatically used because the quantum fields are on-shell.
From now on we abandon the super-script {\it quant} because it will be obvious from the context if we refer 
to the classical expression (\ref{Tint}) or to its quantum counterpart.

In (\ref{2-massless-vector}) we are using the Pauli-Jordan distribution
\be
D_{m}(x) = D_{m}^{(+)}(x) + D_{m}^{(-)}(x)
\ee
where
\be
D_{m}^{(\pm)}(x) =
\pm {i \over (2\pi)^{3}}~\int dp e^{- i p\cdot x} \theta(\pm p_{0}) \delta(p^{2} -
m^{2})
\ee
and
\be
D^{(-)}(x) = - D^{(+)}(- x).
\ee

We comment on the factor from the definitions above; it is in fact non-arbitrary. With this choice we have the following identity
which will be very useful in the following; we have
\begin{lemma}
Suppose that 
$
\xi
$
is verifying the Klein-Gordon equation (\ref{KG})
\be
(\square + m^{2})\xi = 0. 
\ee
Then the following identity is true:
\be
\int d{\bf y} D_{m}(y - x) \stackrel{\leftrightarrow}\partial^{y}_{0} \xi(y) = - \xi(x)
\ee
where we adopt the convention:
\be
A \stackrel{\leftrightarrow}\partial^{y}_{\mu} B = A \partial^{y}_{\mu}B - B \partial^{y}_{\mu}A.
\ee
From here, applying derivatives with respect to $x$ we obtain:
\bea
\int d{\bf y} \partial_{\rho}D_{m}(y - x) \stackrel{\leftrightarrow}\partial^{y}_{0} \xi(y) = \partial_{\rho}\xi(x)
\nonumber\\
\int d{\bf y} \partial_{\rho}\partial_{\sigma}D_{m}(y - x) \stackrel{\leftrightarrow}\partial^{y}_{0} \xi(y) 
= - \partial_{\rho}\partial_{\sigma}\xi(x)
\eea
etc. 
\label{stack}
\end{lemma}
Now, we will consider only the case
$
m = 0
$
and using the previous lemma we have an well-known formula for the so-called {\it BRST current}:
\be
j^{\mu} \equiv : \partial \cdot v_{a} \stackrel{\leftrightarrow}\partial^{\mu} u_{a}:
\ee
Then we have the conservation law: 
\be
\partial_{\mu}j^{\mu} = 0
\ee
and so the expression
\be
Q \equiv \int d{\bf x}~j^{0}(x)
\ee
it is independent of
$
x^{0}.
$
By direct computation we have:
\bea
[j^{\mu}(y), v_{a}^{\nu}(x)] = i~\partial^{\nu}D_{0}(y - x) \stackrel{\leftrightarrow}\partial^{\mu} u_{a}(y)
\nonumber\\
~[j^{\mu}(y), u_{a}] = 0
\nonumber\\
~[j^{\mu}(y), \tilde{u}_{a}(x)] = i~D_{0}(y - x) \stackrel{\leftrightarrow}\partial^{\mu}\partial\cdot v_{a}(y).
\eea

If we integrate over 
$
{\bf y}
$
and apply the previous lemma we obtain the formulas (\ref{Q-vector-null}). 

We can apply now formula (\ref{pqT}) in the trivial case 
$
n = 1
$
and obtain the following commutation relations:
\bea
[ j^{\mu}(y), T(x)] = i~\partial^{\mu}\partial_{\nu}D_{0}(y - x) :\partial\cdot v_{a}(y) B_{a}^{\nu}(x):
- i~\partial_{\nu}D_{0}(y - x) :\partial^{\mu}\partial\cdot v_{a}(y) B_{a}^{\nu}(x):
\nonumber\\
- i \partial^{\mu}\partial_{\nu}D_{0}(y - x) :u_{a}(y) C_{a}^{\nu}(x):
+ i~\partial_{\nu}D_{0}(y -x) \partial^{\mu} :u_{a}(y) C_{a}^{\nu}(x):
\eea
where
\bea
B_{a\nu} \equiv \tilde{u}_{\nu,\rho}\cdot T = - f_{abc} u_{b}v_{c\nu}
\nonumber\\
C_{a\nu} \equiv v_{a\nu}\cdot T = f_{abc} (v^{\rho}_{b}F_{c\rho\nu} - u_{b}\tilde{u}_{c,\nu}).
\label{BC1}
\eea
If we integrate over 
$
{\bf y}
$
and apply the lemma \ref{stack} we obtain the formula 
\be
[Q, T(x)] = i~\partial_{\mu}T^{\mu}
\ee
where
\be
T^{\mu} \equiv f_{abc}~\left( u_{a}~v_{b\nu}~F^{\nu\mu}_{c} - {1\over 2}~f_{abc}~u_{a}~u_{b}~\partial^{\mu}\tilde{u}_{c}\right)
\ee

We can iterate the procedure and derive:
\bea
[ j^{\mu}(y), T^{\nu}(x)] = - i~\partial^{\mu}\partial_{\rho}D_{0}(y - x) :\partial\cdot v_{a}(y) B_{a}^{\nu\rho}(x):
\nonumber\\
+ i~\partial_{\rho}D_{0}(y - x) :\partial^{\mu}\partial\cdot v_{a}(y) B_{a}^{\nu\rho}(x):
\nonumber\\
- i \partial^{\mu}\partial_{\rho}D_{0}(y - x) :u_{a}(y) C_{a}^{\nu\rho}(x):
+ i~\partial_{\rho}D_{0}(y - x) :\partial^{\mu}u_{a}(y) C_{a}^{\nu\rho}(x):
\eea
where
\bea
B_{a\nu\rho} \equiv \tilde{u}_{\nu,\rho}\cdot T_{\nu} = \eta_{\nu\rho}~B_{a}, \qquad B_{a} \equiv {1\over 2}~f_{abc} u_{b}u_{c}
\nonumber\\
C_{a\nu\rho} \equiv v_{a\rho}\cdot T_{\nu} = - f_{abc} u_{b}F_{c\nu\rho}.
\label{BC2}
\eea
If we integrate over 
$
{\bf y}
$
and apply the lemma \ref{stack} we obtain the formula 
\be
[Q, T^{\nu}(x)] = i~\partial_{\rho}T^{\nu\rho}
\ee
where
\be
T^{\nu\rho} \equiv  {1\over 2}~f_{abc}~u_{a}~u_{b}~F_{c}^{\nu\rho}
\ee

Finally it is easy to derive
\be
[ j^{\mu}(y), T^{\nu\rho}(x)] = 0 \qquad \Rightarrow \qquad [Q,  T^{\nu\rho}(x)] = 0.
\ee

The next step is to extend these commutation rules to chronological products. We have:
\begin{thm}
The chronological products can be chosen such that they verify:
\bea
[ j^{\mu}(y), T(T(x_{1}),\dots,T(x_{n}))] = 
\nonumber\\
- i [ ~\partial^{\mu}\partial\cdot v_{a}^{(+)}(y) \sum_{m=1}^{n}~ \partial_{\rho}D_{0}(y - x_{m})~
T(T(x_{1}),\dots,B_{a}^{\rho}(x_{m}),\dots,T(x_{n}))
\nonumber\\
- \partial\cdot v_{a}^{(+)}(y) \sum_{m=1}^{n}~ \partial^{\mu}\partial_{\rho}D_{0}(y - x_{m})~
T(T(x_{1}),\dots,B_{a}^{\rho}(x_{m}),\dots,T(x_{n})) ]
\nonumber\\
+ i\Bigl[ \sum_{m=1}^{n}~ \partial_{\rho}D_{0}(y - x_{m})~
T(T(x_{1}),\dots,B_{a}^{\rho}(x_{m}),\dots,T(x_{n}))~\partial^{\mu}\partial\cdot v_{a}^{(-)}(y)
\nonumber\\
- \sum_{m=1}^{n}~ \partial^{\mu}\partial_{\rho}D_{0}(y - x_{m})~
T(T(x_{1}),\dots,B_{a}^{\rho}(x_{m}),\dots,T(x_{n}))~\partial\cdot v_{a}^{(-)}(y) \Bigl]
\nonumber\\
+ i \Bigl[ ~\partial^{\mu}u_{a}^{(+)}(y) \sum_{m=1}^{n}~ \partial_{\rho}D_{0}(y - x_{m})~
T(T(x_{1}),\dots,C_{a}^{\rho}(x_{m}),\dots,T(x_{n}))
\nonumber\\
- u_{a}^{(+)}(y) \sum_{m=1}^{n}~ \partial^{\mu}\partial_{\rho}D_{0}(y - x_{m})~
T(T(x_{1}),\dots,C_{a}^{\rho}(x_{m}),\dots,T(x_{n})) \Bigl]
\nonumber\\
+ i\Bigl[ \sum_{m=1}^{n}~ \partial_{\rho}D_{0}(y - x_{m})~
T(T(x_{1}),\dots,C_{a}^{\rho}(x_{m}),\dots,T(x_{n}))~\partial^{\mu}u_{a}^{(-)}(y)
\nonumber\\
- \sum_{m=1}^{n}~ \partial^{\mu}\partial_{\rho}D_{0}(y - x_{m})~
T(T(x_{1}),\dots,C_{a}^{\rho}(x_{m}),\dots,T(x_{n}))~u_{a}^{(-)}(y) \Bigl]
\nonumber\\
+ \sum_{k < m}~\sum_{a}~\Bigl[ \partial_{\rho}D_{0}^{(+)}(y - x_{m})~\partial^{\mu}\partial_{\lambda}D_{0}^{(+)}(y - x_{k})
- \partial_{\rho}D_{0}^{(-)}(y - x_{m})~\partial^{\mu}\partial_{\lambda}D_{0}^{(-)}(y - x_{k})
\nonumber\\
- \partial^{\mu}\partial_{\rho}D_{0}^{(+)}(y - x_{m})~\partial_{\lambda}D_{0}^{(+)}(y - x_{k})
+ \partial^{\mu}\partial_{\rho}D_{0}^{(-)}(y - x_{m})~\partial_{\lambda}D_{0}^{(-)}(y - x_{k}) \Bigl]
\nonumber\\
T(T(x_{1}),\dots,B_{a}^{\lambda}(x_{m}),\cdots,C_{a}^{\rho}(x_{k}),\cdots,\dots,T(x_{n}))
\nonumber\\
+ \sum_{k > m}~\sum_{a}~\Bigl[ \partial_{\rho}D_{0}^{(+)}(y - x_{m})~\partial^{\mu}\partial_{\lambda}D_{0}^{(+)}(y - x_{k})
- \partial_{\rho}D_{0}^{(-)}(y - x_{m})~\partial^{\mu}\partial_{\lambda}D_{0}^{(-)}(y - x_{k})
\nonumber\\
- \partial^{\mu}\partial_{\rho}D_{0}^{(+)}(y - x_{m})~\partial_{\lambda}D_{0}^{(+)}(y - x_{k})
+ \partial^{\mu}\partial_{\rho}D_{0}^{(-)}(y - x_{m})~\partial_{\lambda}D_{0}^{(-)}(y - x_{k}) \Bigl]
\nonumber\\
T(T(x_{1}),\dots,C_{a}^{\rho}(x_{m}),\cdots,B_{a}^{\lambda}(x_{k}),\cdots,\dots,T(x_{n}))
\label{jT}
\eea
where the expressions
$
B_{a}^{\lambda},C_{a}^{\rho}
$
are given by (\ref{BC1}).
\end{thm}
The proof is a straightforward application of the theorem \ref{dcomm}. 

To obtain the commutator of $Q$ with the chronological products we have to integrate over 
$
{\bf y}.
$
To do this operation we need the following result which is of the same nature as Lemma \ref{stack}:
\begin{lemma}
The following formula is true:
\be
\int d{\bf y} \partial_{\rho}D_{0}^{(\epsilon)}(y - x_{1})~\stackrel{\leftrightarrow}\partial^{0}
\partial_{\lambda}D_{0}^{(\epsilon)}(y - x_{2}) = 0.
\ee
\end{lemma}
We now have the end result:
\begin{thm}
In the preceding conditions the following formula is true:
\bea
[ Q, T(T(x_{1}),\dots,T(x_{n}))] = 
\nonumber\\
i \sum_{m=1}^{n}~[ - \partial_{\mu}\partial\cdot v_{a}^{(+)}(x_{m}) T(T(x_{1}),\dots,B_{a}^{\mu}(x_{m}),\dots,T(x_{n}))
\nonumber\\
- T(T(x_{1}),\dots,B_{a}^{\mu}(x_{m}),\dots,T(x_{n}))~\partial_{\mu}\partial\cdot v_{a}^{(-)}(x_{m}) 
\nonumber\\
+ \partial_{\mu}u_{a}^{(+)}(x_{m}) T(T(x_{1}),\dots,C_{a}^{\mu}(x_{m}),\dots,T(x_{n}))
\nonumber\\
- T(T(x_{1}),\dots,C_{a}^{\mu}(x_{m}),\dots,T(x_{n}))~\partial_{\mu}u_{a}^{(-)}(x_{m}) ]
\eea
\end{thm}
{\bf Proof:} We do the integration over 
$
{\bf y}
$
and apply the preceding lemma, so the last two terms from the formula (\ref{jT}) give a null contribution. For the first four terms we use 
lemma \ref{stack} for convenient $\phi$ expressions:
\bea
\phi_{1}(y) = \partial\cdot v_{a}^{(+)}(y)~T(T(x_{1}),\dots,B_{a}^{\rho}(x_{m}),\dots,T(x_{n}))
\nonumber\\
\phi_{2}(y) = T(T(x_{1}),\dots,B_{a}^{\rho}(x_{m}),\dots,T(x_{n}))~\partial\cdot v_{a}^{(-)}(y)
\nonumber\\
\phi_{3}(y) = u_{a}^{(+)}(y)~T(T(x_{1}),\dots,B_{a}^{\rho}(x_{m}),\dots,T(x_{n}))
\nonumber\\
\phi_{4}(y) = T(T(x_{1}),\dots,B_{a}^{\rho}(x_{m}),\dots,T(x_{n}))~u_{a}^{(-)}(y).
\nonumber
\eea
$\qed$

Similar formulas can be proven for other chronological products, for instance:
\bea
[ Q, T(T^{\mu}(x_{1}),T(x_{2}),\dots,T(x_{n}))] = 
\nonumber\\
i [ \partial^{\mu}\partial\cdot v_{a}^{(+)}(x_{m}) T(B_{a}(x_{1}),T(x_{2}),\dots,T(x_{n}))
\nonumber\\
+ T(B_{a}(x_{1}),T(x_{2}),\dots,T(x_{n}))~\partial^{\mu}\partial\cdot v_{a}^{(-)}(x_{1}) 
\nonumber\\
+ \partial_{\nu}u_{a}^{(+)}(x_{1}) T(C_{a}^{\nu\mu}(x_{1}),T(x_{2}),\dots,T(x_{n}))
\nonumber\\
+ T(C_{a}^{\nu\mu}(x_{1}),T(x_{2}),\dots,T(x_{n}))~\partial_{\nu}u_{a}^{(-)}(x_{1}) ]
\nonumber\\
- i \sum_{m=2}^{n}~[ \partial_{\nu}\partial\cdot v_{a}^{(+)}(x_{m}) T(T^{\mu}(x_{1}),T(x_{2}),\dots,B_{a}^{\nu}(x_{m}),\dots,T(x_{n}))
\nonumber\\
+ T(T^{\mu}(x_{1}), T(x_{2}),\dots,B_{a}^{\nu}(x_{m}),\dots,T(x_{n}))~\partial_{\nu}\partial\cdot v_{a}^{(-)}(x_{m}) 
\nonumber\\
+ \partial_{\nu}u_{a}^{(+)}(x_{m}) T(T^{\mu}(x_{1}),T(x_{2}),\dots,C_{a}^{\nu}(x_{m}),\dots,T(x_{n}))
\nonumber\\
+ T(T^{\mu}(x_{1}),T(x_{2}),\dots,C_{a}^{\nu}(x_{m}),\dots,T(x_{n}))~\partial_{\mu}u_{a}^{(-)}(x_{m}) ].
\eea
\newpage
 
\section{Conclusions}
We note that the commutators of the type
$
[ Q, T(T^{I_{1}}(x_{1},\dots,\dots,T^{I_{n}}(x_{n})) ]
$
are given by sums of terms where a factorization property appears: the factors
$
\partial_{\mu}\partial\cdot v_{a}, \partial_{\mu}u_{a} 
$
are pulled out and we remain with chronological products where one of the entry is a Wick submonomial i.e. a derivative of the type
(\ref{derivative}). 

On the other hand we know that the gauge invariance condition in order $n$ is can be written using the (linear) BRST operator 
\be
s \equiv d_{Q} - i~\delta
\ee
where
\be
d_{Q}A \equiv [Q,A]
\ee
and
\be
(\delta C)^{I_{1},\dots,I_{n}}
= \sum_{l=1}^{n} (-1)^{s_{l}} {\partial\over \partial x^{\mu}_{l}}
C^{I_{1},\dots,I_{l}\mu,\dots,I_{n}}.
\ee

Then, the on shell gauge invariance is:
\be
sT(T^{\mu}(x_{1}),T(x_{2}),\dots,T(x_{n})) = 0
\label{gauge-n}
\ee

One can prove that such an identity is true up to order three \cite{third order}, but a general proof in all orders is missing. If
we inspect the expression of the commutator
$
[ Q, T(T(x_{1}),\dots,\dots,T(x_{n})) ]
$
given by the preceding theorem we notice an interesting point, namely the gauge charge is a sum 
\be
Q = \sum_{m=1}^{n}~Q_{m}
\ee
where
\bea
[Q_{1}, T(T(x_{1}),\dots,T(x_{n}))] = 
\nonumber\\
i [ - \partial_{\mu}\partial\cdot v_{a}^{(+)}(x_{1}) T(B_{a}^{\mu}(x_{1}),T(x_{2}),\dots,T(x_{n}))
\nonumber\\
- T(B_{a}^{\mu}(x_{1}),T(x_{2}),\dots,T(x_{n}))~\partial_{\mu}\partial\cdot v_{a}^{(-)}(x_{1}) 
\nonumber\\
+ \partial_{\mu}u_{a}^{(+)}(x_{1}) T(C_{a}^{\mu}(x_{1}),T(x_{2}),\dots,T(x_{n}))
\nonumber\\
- T(C_{a}^{\mu}(x_{1}),T(x_{2}),\dots,T(x_{n}))~\partial_{\mu}u_{a}^{(-)}(x_{1}) ]
\eea
and
$
Q_{m}
$
can be obtained from
$
Q_{1}
$
making
$
1 \leftrightarrow m.
$
Such a formula is not obvious from the general definition. One is tempted to conjecture that the operator
$\delta$
must also have such a form. This seems to be plausible because similar operators, as for instance, the commutator of the translation operator
$
P_{\mu}
$
with various chronological products can be written is a similar form. The reason is that 
$
P_{\mu}
$
is an expression bilinear in the (quantum) fields so we can apply the preceding arguments in this case also. However, the operator
$\delta$
is more complicated and we did not succed to prove such a formula. However, we notice that it might be a good idea to reduce the 
gauge invariance problem (\ref{gauge-n}) to ``simpler" problems where we have Wick submonomials as entries. This idea first appeared
in \cite{algebra} and it is worthwile to study further.

\newpage

{\bf Acknowledgments:} The author acknowledges the financial support received from the Romanian
Ministry of Research, Digitalisation and Innovation through the Project PN 19 06 01 01/2019.


\end{document}